\documentclass[%
 reprint,
 amsmath,amssymb,
 aps
pra,
]{revtex4-1}
\usepackage{graphicx}
\usepackage{dcolumn}
\usepackage{bm}

\usepackage{braket}
\def\ra{\rangle}

\def\hbs{\vert\psi(\alpha,m)\ra}
\def\bcen{\begin{center}}
\def\ecen{\end{center}}
\usepackage{epstopdf}
\begin{document}

\preprint{APS/123-QED}

\title{Number state filtered coherent state}

\author{Nilakantha Meher}
\email{nilakantha.meher6@gmail.com}
\author{S. Sivakumar}
 \email{siva@igcar.gov.in}
\affiliation{Materials Physics Division, Indira Gandhi Centre For Atomic Research,
Homi Bhabha National Institute, Kalpakkam, India-603102}
\begin{abstract}
Number state filtering in coherent states leads to  sub-Poissonian photon statistics. These states are more suitable for phase estimation when compared with the coherent states. Nonclassicality of these states is quantified in terms of the negativity of the Wigner function and the entanglement potential.   Filtering of the vacuum from a coherent state is almost like the photon-addition.  However,  filtering makes the state more resilient against dissipation than photon-addition.  Vacuum state filtered coherent states perform better than the photon-added coherent states for  a two-way quantum key distribution protocol. A scheme to generate these states in multi-photon atom-field interaction  is presented. 
\end{abstract}
\maketitle
\section{Introduction}
Dynamics of the classical electromagnetic field  is described by the Maxwell equations.  All the features describable within this formalism are classified as the classical aspects of the field. However, this description is inadequate as there are experimental observations such as the anti-bunching of light, quadrature squeezing, sub-Poissonian photon statistics, etc.,  which cannot be understood within the classical description \cite{Gerry}.  These observations are explainable in the quantum description of the electromagnetic field.  In this formulation, the most fundamental states of the electromagnetic field are the number states, which correspond to states of definite number of photons.   It is of  interest to know those quantum states of the field which have features similar to the classical field.  It turns out that the coherent state, which is a superposition of all the number states with specific superposition coefficients, has that property \cite{Glau, Sudar}. This specific choice of  the superposition coefficients makes the coherent state very different from these number states.  For instance, the coherent states are  the eigenstates of the annihilation operator $\hat{a}$, the corresponding complex eigenvalue $\alpha$ is the coherent state amplitude. These states form an over-complete basis set so that any state of the electromagnetic field is expressible as a continuous superposition. As a consequence, the expectation values of an operator-valued function of the annihilation operator is the statistical average of the corresponding function of the coherent state amplitude.   The expression for the coherent state in  the number state basis is 
\begin{align}\label{Coherentstate}
\ket{\alpha}=e^{-|\alpha|^2/2}\sum_{n=0}^\infty \frac{\alpha^n}{\sqrt{n!}}\ket{n}.
\end{align}

 Correlation functions of any order are factorizable for the coherent states \cite{Glau2}.  For large amplitude coherent state, the photon number fluctuation is much smaller than the mean number of photon. Hence large amplitude coherent states represent the classical field.  These states saturate the  uncertainties in the field quadratures to their minimum values allowed by quantum mechanics. This, in turn, implies that the  phase space distribution of a coherent state  is well localized. In the context of a harmonic oscillator, the coherent state wave packet oscillates without dispersion. An alternate definition identifies  the  coherent state as a displaced vacuum state \cite{Gerry}. Interestingly, one of the quasi-probability distributions, namely, the Glauber-Sudarshan P-function for the coherent state is a Dirac delta-function, which is a well defined probability density. For this reason these states are called quasi-classical states.  Any other choice of coefficients makes the resultant state non-classical whose P-function is more singular than a delta function. This signals the nonclassical aspect of the quantum state \cite{ Glau,Sudar}. 

Nonclassicality of quantum states is of importance in quantum optics \cite{Mand,Anirban}. The notion of  nonclassicality is based on the quasi-probability distributions for the quantum states. Negative value of P-function or Wigner function is sufficient to identify a quantum state as nonclassical. 
Many interesting nonclassical states have been defined  by suitable modifications of the quasi-classical coherent states; for instance,  photon-added coherent states \cite{Agar}, even/odd coherent states \cite{Dodo2}, truncated coherent states \cite{Miranowicz,Siva3}, etc. The even and odd coherent states are defined as the symmetric and anti-symmetric superposition of two coherent states whose complex  amplitudes are of equal magnitude but out of phase  by $\pi$ \cite{Dodo2}.  These states exhibit quadrature squeezing and sub-Poissonian statistics for suitable values of the amplitude. Agarwal and  Tara introduced a new class of nonclassical states by adding quanta to the  coherent state. This states are called photon added coherent state (PACS) \cite{Agar}.  For proper choice of parameters, PACS  shows quadrature squeezing and sub-Poissonian photon statistics. Truncated coherent states are obtained by modifying the expression in Eqn. \ref{Coherentstate} by removing either a finite number of contiguous states starting from the vacuum or by removing all the states beyond a specified number state \cite{Miranowicz,Siva3}.  Another important class of nonclassical states is  obtained by number state hole burning or filtering in the coherent states \cite{Baseia}.   It is known that any pure state in which the  vacuum state $\ket{0}$ is absent is nonclassical \cite{Lee,Lima}. 
A possible generalization is to consider states in which a number state, not necessarily the vacuum, is absent.  More generally, states with many of the number states not being present in the superposition have been studied to assess their  possible use in  practical applications such as optical data storage and optical communication \cite{Baseia, Baseia2}.   It may be noted that schemes to generate such states  are known\cite{ Sanaka, Malb, Avelar,Avelar2, GerryAdil, Zaguary, Himel, Escher, Merlin, Zou}.\\

In this paper, we study the non classical properties of number state filtered coherent state. The photon counting statistics of these states is discussed in Section \ref{HBCS}. Sub-Poissonian statistics is shown to improve phase estimation in comparison to the coherent states. In Section \ref{Nonclassical}, quantification of nonclassicality of these states is carried out. Entanglement potential and negativity of Wigner function are studied in this context. In Section \ref{Generation}, the possibility of generating these states in the interaction of a 3-level atom with a cavity field is discussed. In Section \ref{QIP}, features of vacuum state filtered coherent state are analyzed. Specifically, the robustness of the state against dissipation is exhibited. This feature enables these states to perform better than the single photon added coherent state (SPACS) when it comes to two-way quantum key distribution. Use of these states as a source to generate the even and odd coherent states is discussed in Section \ref{Catstate}.

\section{Number state filtering in coherent state}\label{HBCS}
  Number state filtered state (NSFS), denoted by  $\ket{\psi(\alpha,m)}$, is defined as 
\begin{align}\label{HBcoherentstate}
\ket{\psi(\alpha,m)}=\frac{e^{-|\alpha|^2/2}}{N_m}\sum_{n=0,n\ne m}^\infty \frac{\alpha^n}{\sqrt{n!}}\ket{n},
\end{align}
where $N_m$ is the normalization constant given by
\begin{align}
N_m=\sqrt{1-e^{-|\alpha|^2}\frac{|\alpha|^{2m}}{m!}}.
\end{align}
This definition implies that the state is obtained if the number state $\ket{m}$ is absent in the superposition defined in Eqn. \ref{Coherentstate}.\\

 The state $\ket{\psi(\alpha,m)}$ does not  become $\ket{\alpha}$ for any $m$. However, if $m>>|\alpha|$ or vice-versa, then $\hbs \approx \ket{\alpha}$.  It is worth noting that the coherent state is a superposition of NSFSs, 
\begin{align}
\ket{\alpha}=\frac{N_m}{2}(\ket{\psi(\alpha,m)}&+\ket{\psi(-\alpha,m)})\nonumber\\
&+\frac{N_k}{2}(\ket{\psi(\alpha,k)}-\ket{\psi(-\alpha,k)}),
\end{align}
where $m$ is odd and $k$ is even.\\ 
NSFS is expressible as
\begin{align}
\ket{\psi(\alpha,m)}&=\frac{1}{N_m}(\ket{\alpha}-C_m\ket{m})=\frac{1}{N_m}(I-\ket{m}\bra{m})\ket{\alpha},
\end{align}
a superposition of  $\ket{\alpha}$ and  $\ket{m}$, where $C_m=e^{-|\alpha|^2/2}\frac{\alpha^m}{\sqrt{m!}}$ \cite{Baseia3}. The overlap between the states $\hbs$ and $\ket{\alpha}$ is 
\begin{align}
|\langle \alpha|\psi(\alpha,m)\rangle|^2=1-|C_m|^2,
\end{align}
which is minimum if $|\alpha|^2=m$. Therefore, maximum nonclassicality of $\hbs$ is expected if $|\alpha|^2=m$.\\

For a given $\alpha$, the state  $\ket{\psi(\alpha,m)}$ corresponding to different $m$ are linearly independent. Hence the set $\{\ket{\psi(\alpha,m)}\}_{m=0}^\infty$ can be used as a basis. Further, for a given $m$ 
\begin{align}
\frac{1}{\pi}\int N_m^2 \ket{\psi(\alpha,m)}\bra{\psi(\alpha,m)} d^2\alpha=I-\ket{m}\bra{m},
\end{align}
which is the completeness relation for the Hilbert space less the number state $\ket{m}$ of a harmonic oscillator.\\
It is interesting to note that the even and odd coherent states are expressible as 
\begin{align}
\ket{ECS}=\frac{1}{\sqrt{2}}\sqrt{\frac{1-|C_m|^2}{1+|C_0|^4}}(\ket{\psi(\alpha,m)}&+\ket{\psi(-\alpha,m)}),\nonumber\\&( \text{for odd} ~m)
\end{align}
and
\begin{align}
\ket{OCS}=\frac{1}{\sqrt{2}}\sqrt{\frac{1-|C_m|^2}{1-|C_0|^4}}(\ket{\psi(\alpha,m)}&-\ket{\psi(-\alpha,m)}),\nonumber\\&( \text{for even} ~m).
\end{align}

\subsection{Photon Statistics}\label{photonstatistics}
\begin{figure}[h]
\centering
\includegraphics[width=9.5cm, height=4.5cm]{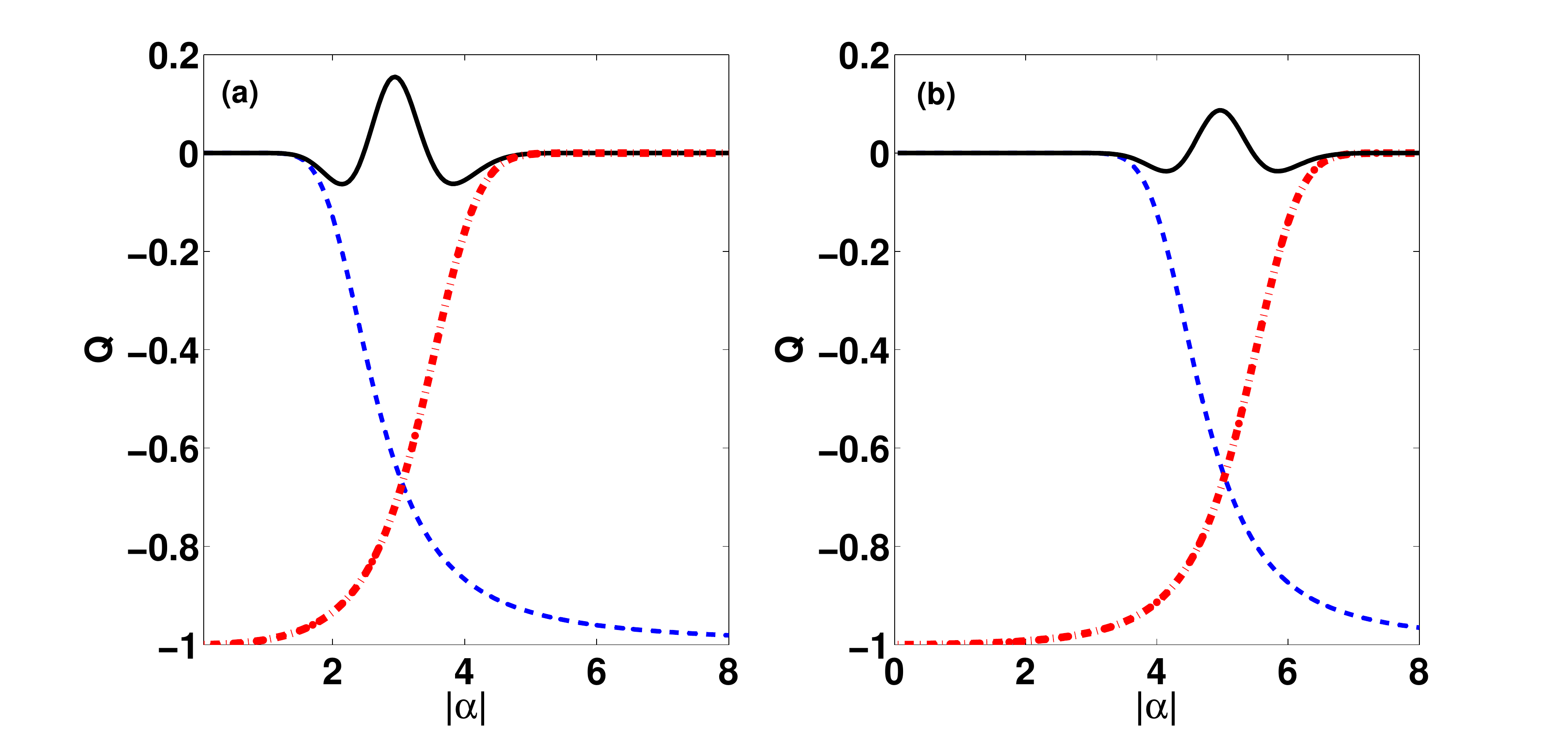}
\caption{The $Q$ parameter as a function of $|\alpha|$ for the state $\ket{\psi(\alpha,m)}$ (continuous), $\ket{\alpha,m-1}_u$ (dashed) and $\ket{\alpha,m}_l$ (dot-dashed) with $(a) m=9$ and $(b)m=25$.}
\label{Qforall}
\end{figure}

Photon statistics is an experimentally measurable aspect of the electromagnetic field. Probability of detecting $k$ photons in a measurement is $|\langle k|\psi(\alpha,m)\rangle|^2$, where $\ket{k}$ represents the number state with $k$ photons. Using the number state expression given in Eqn. \ref{HBcoherentstate},
\begin{align}
P_k=|\langle k|\psi(\alpha,m)\rangle|^2=\frac{e^{-|\alpha|^2}}{N_m^2}\left|\frac{\alpha^k}{\sqrt{k!}}-\frac{\alpha^m}{\sqrt{m!}}\delta_{k,m}\right|^2,
\end{align}
which gives the average number of photons in the state as
\begin{align}\label{AvgPhoton}
\langle \hat{n}\rangle=\langle a^\dagger a \rangle=\frac{1}{N_m^2}\left(|\alpha|^2-m e^{-|\alpha|^2}\frac{|\alpha|^{2m}}{m!}\right).
\end{align}
For $|\alpha|>> m$ or  $|\alpha| << m$ 
\begin{align}
\langle \hat{n}\rangle\approx |\alpha|^2,
\end{align}
as expected since $\hbs\approx\ket{\alpha}$ under these limits.
Surprisingly, if $|\alpha|^2=m$,
\begin{align}
\langle \hat{n}\rangle= |\alpha|^2,
\end{align}
 same as the average number of photons in the coherent state $\ket{\alpha}$ even though overlap between $\hbs$ and $\ket{\alpha}$ is minimum. Nevertheless, the nonclassical nature of the photon distribution can be identified by the Mandel Q-parameter
\begin{align}
Q=\frac{\langle a^{\dagger 2}a^2\rangle-\langle a^\dagger a\rangle^2}{\langle a^\dagger a\rangle}.
\end{align}
If $Q=0$, the state is Poissonian which corresponds to a coherent state. If $Q<0$, the state is sub-Poissonian which is a nonclassical feature. The state is super-Poissonian if $Q>0$. For the state $\ket{\psi(\alpha,m)}$,
\begin{align}
Q=\frac{e^{-|\alpha|^2}|\alpha|^{2m}}{m!N_m^2 \langle\hat{n}\rangle}\left(2m|\alpha|^2\right.&-m(m-1)\nonumber\\&\left.-|\alpha|^4-me^{-|\alpha|^2}\frac{|\alpha|^{2m}}{m!}\right),
\end{align}
where $\langle \hat{n}\rangle$ is given in Eqn. \ref{AvgPhoton}.
The dependence of $Q$ on $|\alpha|$ is shown in Fig. \ref{Qforall} $(a)$ for $m=9$ and Fig. \ref{Qforall} $(b)$ for $m=25$. The state is characterized by $Q<0$ as well as $Q>0$ in different ranges of $|\alpha|$.
To understand the photon statistics of  $\hbs$,  consider expressing it as a linear superposition of a lower truncated coherent state (LTCS) and an upper truncated coherent state (UTCS) whose number states expressions are \cite{Miranowicz,Siva3}
\begin{align}
\ket{\alpha,N}_l&=\tilde{N}_l\sum_{n=0}^{\infty}\frac{\alpha^{N+1+n}}{\sqrt{(N+1+n)!}}\ket{N+1+n},
\end{align}
and
\begin{align}
\ket{\alpha,N}_u&=\tilde{N}_u\sum_{n=0}^{N}\frac{\alpha^n}{\sqrt{n!}}\ket{n},
\end{align}  
respectively. Here $\tilde{N}_u^{-2}=e^{|\alpha|^2}(1-\gamma(N+1,|\alpha|^2)/N!)$ and $\tilde{N}_l^{-2}=e^{|\alpha|^2}\gamma(N+1,|\alpha|^2)/N!$ with $\gamma(N,x)=(N-1)!\left[1-e^{-x}\sum_{j=0}^{N-1}\frac{x^j}{j!}\right],$ the incomplete Gamma function. Note that, the states UTCS and LTCS are orthogonal to each other.\\
     
An interesting aspect of these two states is that they are sub-Poissonian for any $\alpha$.  As $\vert\alpha\vert$ increases to values much larger than unity,  the photon number distribution of LTCS  becomes Poissonian, that is, $Q$ tends to zero.  In the same limit, the $Q$ parameter of UTCS becomes -1,  that is, sub-Poissonian. \\ 

The state $\hbs$ is expressible as a linear combination of LTCS and UTCS:
\begin{align}
\hbs=C_u\ket{\alpha,m-1}_u+C_l\ket{\alpha,m}_l,
\end{align}
where $C_u=~{_u}\bra{\alpha,m-1}\psi(\alpha,m)\rangle$ and $C_l=~{_l}\bra{\alpha,m}\psi(\alpha,m)\rangle$.
\begin{figure}[h]
\centering
\includegraphics[width=9cm, height=7cm]{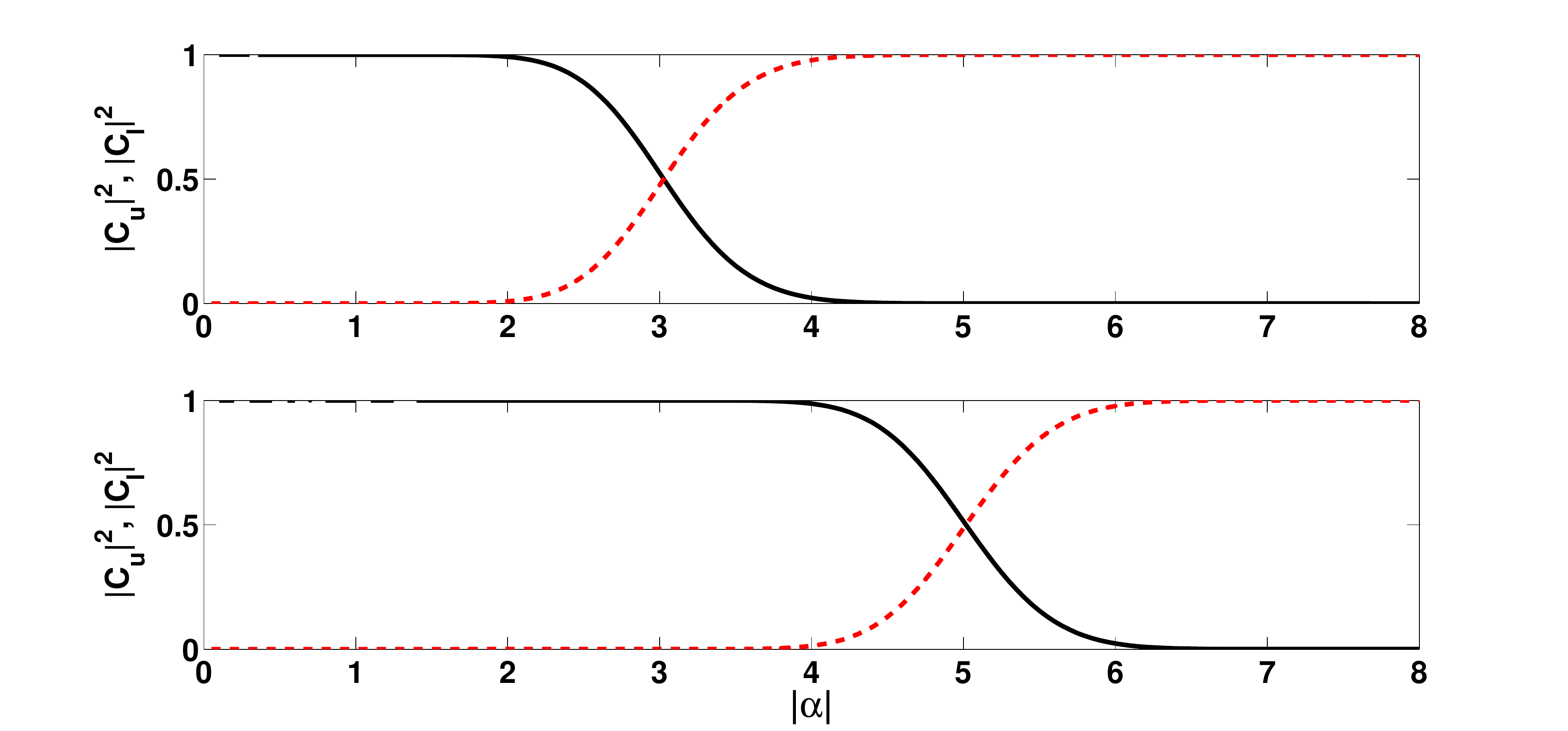}
\caption{ $|C_u|^2$ (continuous)  and $|C_l|^2$ (dashed) are plotted as a function of $|\alpha|$ for $(a)m=9$ and $(b)m=25$.   }
\label{SupCoeff}
\end{figure} \\
Though the states involved in the superposition are sub-Poissonian, the resultant state is super-Poissonian for a range of values of $\alpha$ as shown in  Fig. \ref{Qforall}.   A key observation is that the state $\hbs$ exhibits maximum super-Poissonian character when  UTCS and LTCS are equally sub-Poissonian.   As $\alpha$ changes,  $\hbs$ exhibits sub-Poissonian feature over a small region and becomes nearly Poissonian for larger and smaller  values of $\vert\alpha\vert$.   This feature emerges here in a way that is analogous to the appearance of super-Poissonian statistics when two number states, which are sub-Poissonian,  are superposed.  If the superposition coefficients are nearly equal in magnitude,  measurement of photon number gives results corresponding to the two number states  with nearly equal probability.  That is, the measurement results fluctuate and the statistics  is super-Poissonian.    To relate this with the behaviour of $Q$ parameter of $\hbs$, $|C_u|^2$ and $|C_l|^2$ are shown as a function of $\vert\alpha\vert$ for different values of $m$  in Fig. \ref{SupCoeff}.  When one of the coefficients is nearly unity, $\hbs$ has negative $Q$ corresponding to either LTCS or UTCS both of which are sub-Poissonian.  When the coefficients are nearly equal, the statistics exhibits fluctuations as in the case of superposition of two number states and the resultant state is super-Poissonian.\\

\begin{figure}[h]
\centering
\includegraphics[width=9cm, height=7cm]{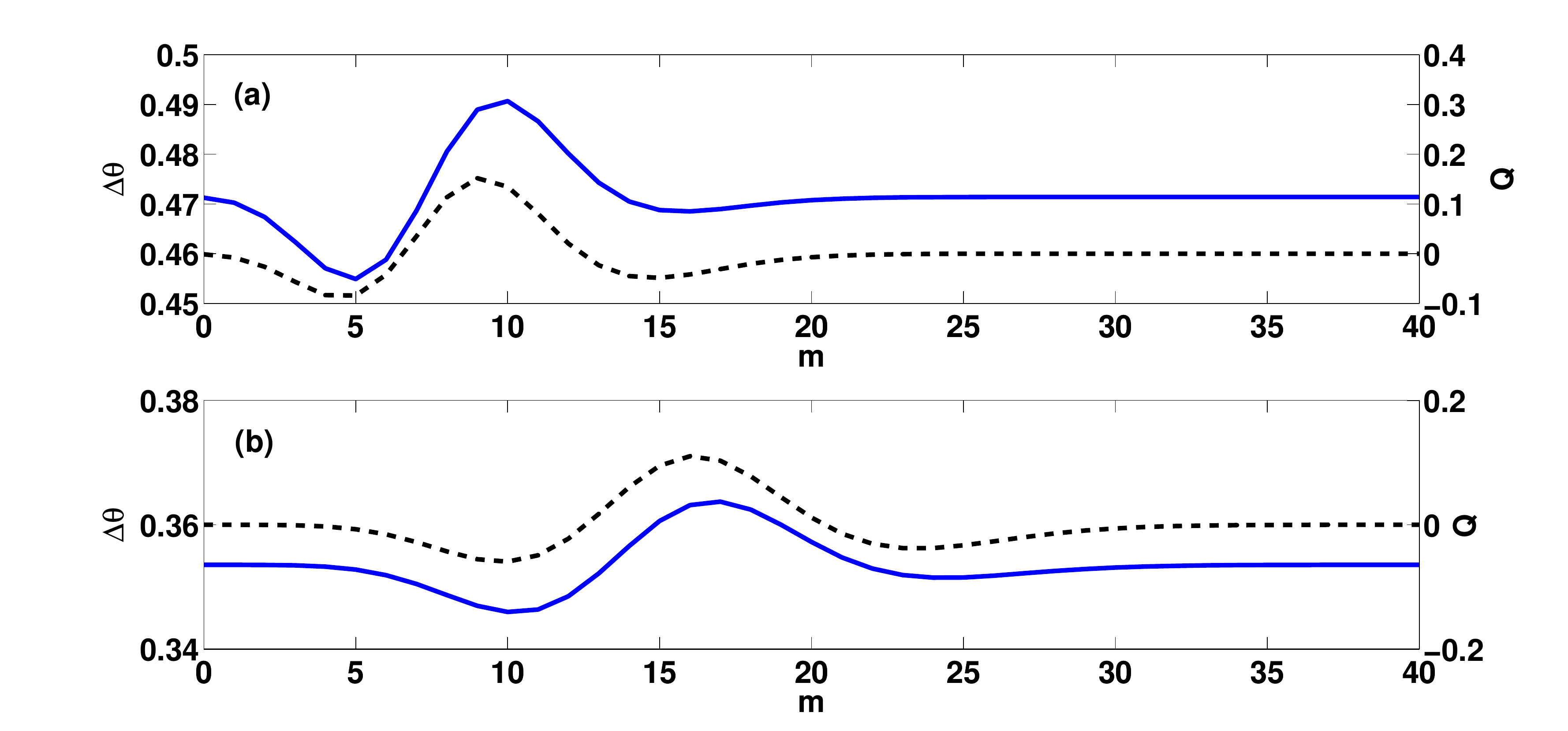}
\caption{ Phase sensitivity $\Delta\theta$ (continuous) and $Q$ parameter (dashed) as a function of $m$ for $\ket{\psi(\alpha,m)}$ for $(a) \alpha=3$ and $(b)\alpha=4$. }
\label{PhaseSensitivity}
\end{figure}

The fact that the states $\hbs$ exhibit sub-Poissonian character indicates that they could be of use in phase estimation using interferometers. Coherent states provide a phase resolution of  $\Delta\theta=1/\sqrt{|\alpha|^2}|\sin\theta|$ \cite{Gerry}.  There is mild improvement if one uses $\hbs$ instead of the coherent states.  This feature is shown in Fig. \ref{PhaseSensitivity},   where the minimum detectable phase is shown as a function of $m$. It is clear that sensitivity is maximum when $\hbs$ exhibits minimum $Q$.  

\subsection{Subtraction of photons from NSFS}
\begin{figure}[h]
\centering
\includegraphics[width=9cm, height=5cm]{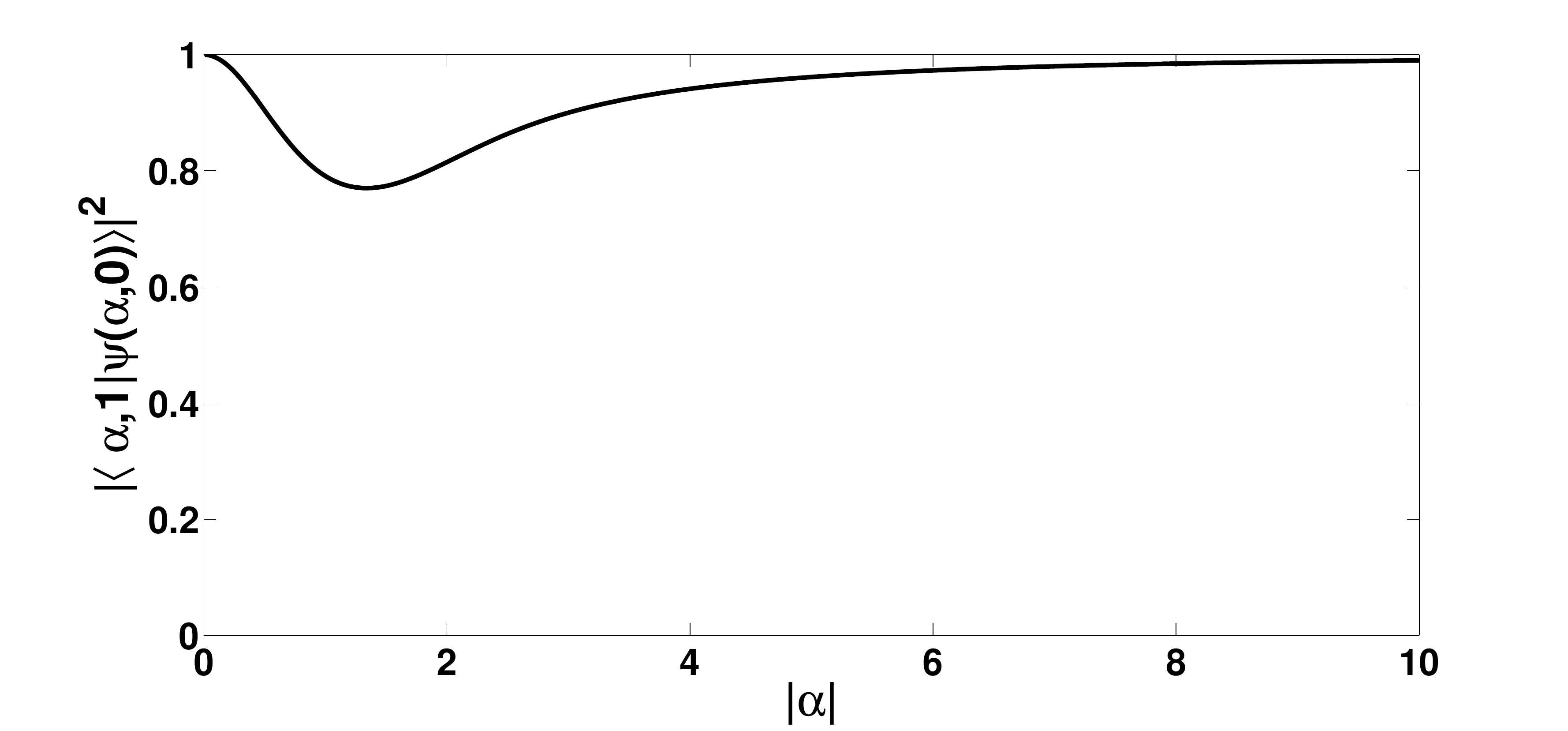}
\caption{Overlap of the state $\ket{\psi(\alpha,0)}$ with single photon added coherent state $\ket{\alpha,1}$ as a function of $|\alpha|$.  }
\label{Fidelity}
\end{figure}

   Photon-addition to a coherent state generates a nonclassical state \cite{Agar}. Like photon addition, photon subtraction is a process which may introduce nonclassicality in many states. However, photon subtraction  leaves the coherent state unaltered as it is  an eigenstate of $\hat{a}$. Action of $\hat{a}$ on other states generates photon-subtracted states \cite{Taka}. One way of realizing photon subtraction experimentally is by  using beam splitters. Though $\ket{\psi(\alpha,m)}$ is necessarily nonclassical, the effect of photon subtraction is of interest. The resultant state on subtracting a photon from $\ket{\psi(\alpha,m)}$ is
\begin{align}
\hat{a}\ket{\psi(\alpha,m)}=\alpha\frac{\sqrt{1-|C_{m-1}|^2}}{{\sqrt{1-|C_m|^2}}}\ket{\psi(\alpha,m-1)},
\end{align}
in which the number state $\ket{m-1}$ is filtered. Successive action of $\hat{a}$ leads to a sequence of NSFS wherein the successively lower number states are filtered. After $m$ photon subtractions, $\ket{\psi(\alpha,m)}$ becomes
\begin{align}
\hat{a}^m\ket{\psi(\alpha,m)}=\alpha^m\frac{\sqrt{1-|C_{0}|^2}}{{\sqrt{1-|C_m|^2}}}\ket{\psi(\alpha,0)},
\end{align}
in which the vacuum state is absent. This state is similar to the SPACS $\ket{\alpha,1}$ \cite{Agar}. The overlap between these two states is 
\begin{align}
|\bra{\alpha,1}\psi(\alpha,0)\rangle|^2=\frac{|\alpha|^2}{(1-e^{-|\alpha|^2})(1+|\alpha|^2)},
\end{align} 
which is plotted as a function of $|\alpha|$ in Fig. \ref{Fidelity}. It clearly indicates that for small or large $|\alpha|$, the states are nearly the same as the overlap is nearly unity.\\

Interestingly, subtraction of one more photon from  $\ket{\psi(\alpha,0)}$ yields
\begin{align}\label{CohfromHBS}
\hat{a}^{m+1}\ket{\psi(\alpha,m)}=\frac{\alpha^{m+1}}{{\sqrt{1-|C_m|^2}}}\ket{\alpha},
\end{align}
the coherent state. In short, subtraction of $m+1$ photons from  $\ket{\psi(\alpha,m)}$ generates the coherent state. This has to be compared with the action of $\hat{a}$ on a number state $\ket{m}$. After $m$ successive photon subtractions, $\ket{m}$ becomes the vacuum state $\ket{0}$ and one more subtraction annihilates the state. 
It is to be noted that while subtraction of $m+1$ photons from $\hbs$ generates a coherent state, addition of photons to a coherent state does not generate $\ket{\psi(\alpha,m)}$, a consequence of the non-commutativity of $a$ and $a^\dagger$.\\

From Eqn. \ref{CohfromHBS}, it is seen that  $\ket{\psi(\alpha,m)}$ satisfies the eigenvalue equation
\begin{align}
\left[(I-\hat{P}_m)a^{m+1}\right]\ket{\psi(\alpha,m)}=\alpha^{m+1}\ket{\psi(\alpha,m)},
\end{align} 
where $\hat{P}_m=\ket{m}\bra{m}$.\\

\begin{figure}[h]
\centering
\includegraphics[width=9cm, height=5cm]{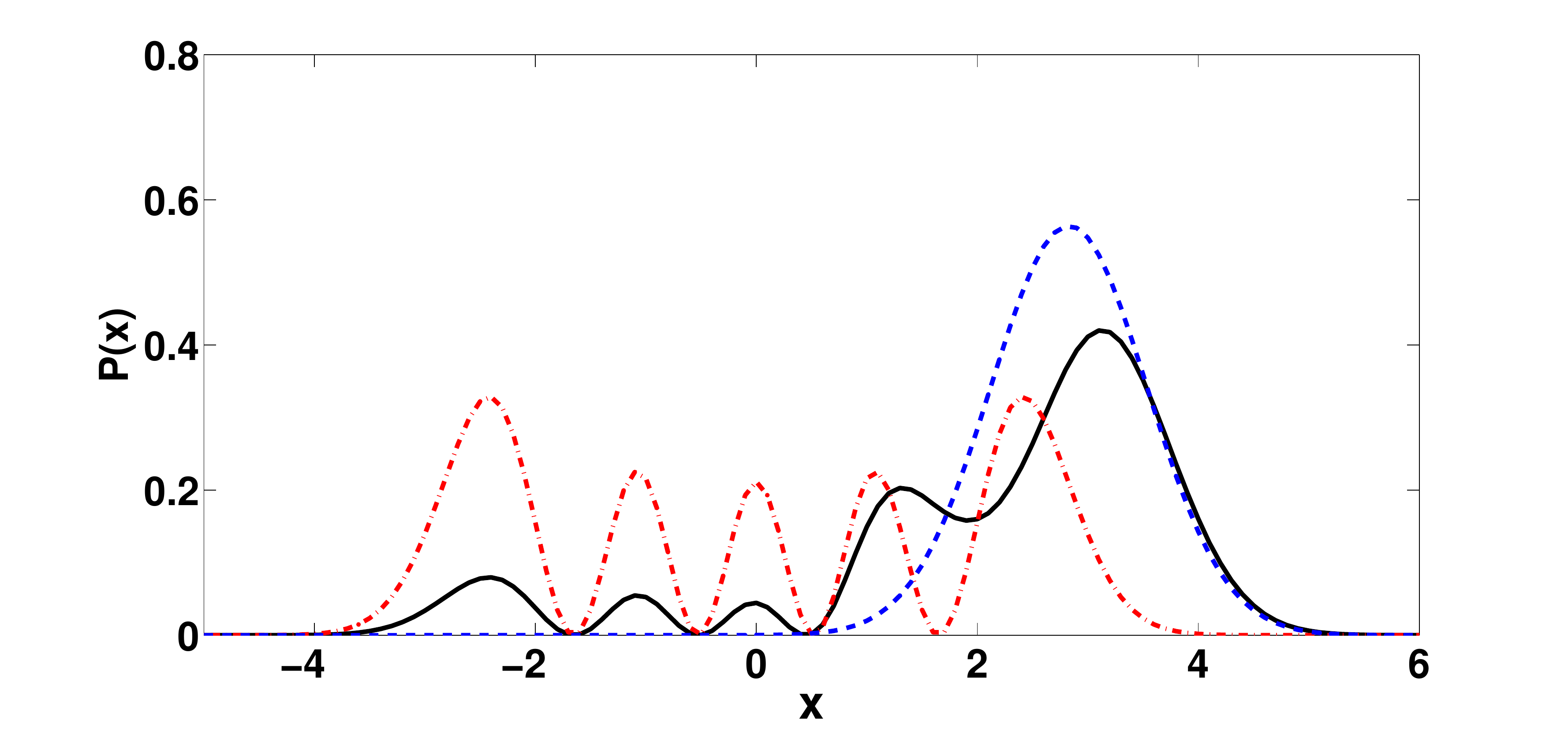}
\caption{Probability distribution for $\ket{x}$ in $\ket{\psi(\alpha,m)}$ (continuous), $\ket{\alpha}$ (dashed) and $\ket{m}$ (dot-dashed) as a function of $x$ for $\alpha=2$ and $m=4$.  }
\label{PositionRep}
\end{figure}

In $x$-representation, the wavefunction for a coherent state is a Gaussian. 
Probability density for $x$ in $\hbs$ is $|\psi_{\alpha,m}(x)|^2$, which is shown in Fig. \ref{PositionRep}. For comparison, the corresponding probability densities for coherent state $\ket{\alpha}$ of amplitude $\alpha=2$ and the number state $\ket{m}$ with $m=4$ is also shown. From the figure it is clear that, the probability distribution is not a Gaussian. In fact, the distribution is oscillatory.

Considering the time evolution of $\ket{\psi(\alpha,m)}$ under $U=e^{-i\omega t a^\dagger a}$, the evolved state is 
\begin{align}
\ket{\psi(\alpha,m,t)}&=\frac{1}{N_m}\left(\ket{\alpha e^{-i\omega t}}-C_m e^{-im\omega t/\hbar}\ket{m}\right),\nonumber\\
&=\ket{\psi(\alpha e^{-i\omega t},m)},
\end{align}
analogous to the evolution of the coherent state $\ket{\alpha} \rightarrow \ket{\alpha e^{-i\omega t}}$.\\

On filtering the number state $\ket{m}$ from the coherent state $\ket{\alpha}$, the resultant state has maximum deviation from $\ket{\alpha}$ if $ |\alpha|^2 \approx m$, which is the condition for maximum nonclassicality of the NSFS. On the other hand, NSFS exhibits maximum super-Poissonian photon statistics under the same condition. In fact, a state with super-Poissonian photon statistics does not refer to the nonclassicality of the state. In order to clarify the condition for maximum nonclassicality of NSFS, we study the negativity of Wigner function and entanglement potential in the next section. 
\section{Nonclassicality of NSFS} \label{Nonclassical}
Phase space representation of quantum states provides additional insights.  For instance, classical aspects of quantum states are readily obtained by such representations. In quantum optics, the most used phase space distributions are the $P$-, $Q$- and Wigner-functions. Negative values of Glauber-Sudarshan $P$-function is a sufficient condition for qualifying the state as non-classical. The condition $\langle (\Delta \hat{n})^2\rangle < \langle \hat{n}\rangle$ requires the $P$-function to be negative. Hence, the amplitude squeezing is a non-classical effect \cite{Gerry}. A state with amplitude squeezing possesses sub-Poissonian photon statistics. The sub-Poissonian character of $\hbs$ has been discussed in the Section \ref{HBCS}.      
\subsection{Wigner function}\label{WignerFunction}
 In this section, we discuss the nonclassicality of $\hbs$ in terms of its Wigner function.  Wigner function, though not singular for any state,  can become negative in some regions.  The  negativity is an indicator of quantumness \cite{Wign}.   Wigner function for a state $\ket{\phi}$ is,
\begin{align}
W(\beta)=\frac{2}{\pi}\sum_{n=0}^\infty (-1)^n\bra{\phi}D(\beta)\ket{n}\bra{n}D^\dagger (\beta)\ket{\phi},
\end{align}
which is an expression in terms of the displaced number state $D(\beta)\ket{n}$ \cite{Gerry, Bishop}.
\begin{figure}[h]
\centering
\includegraphics[width=9cm, height=6cm]{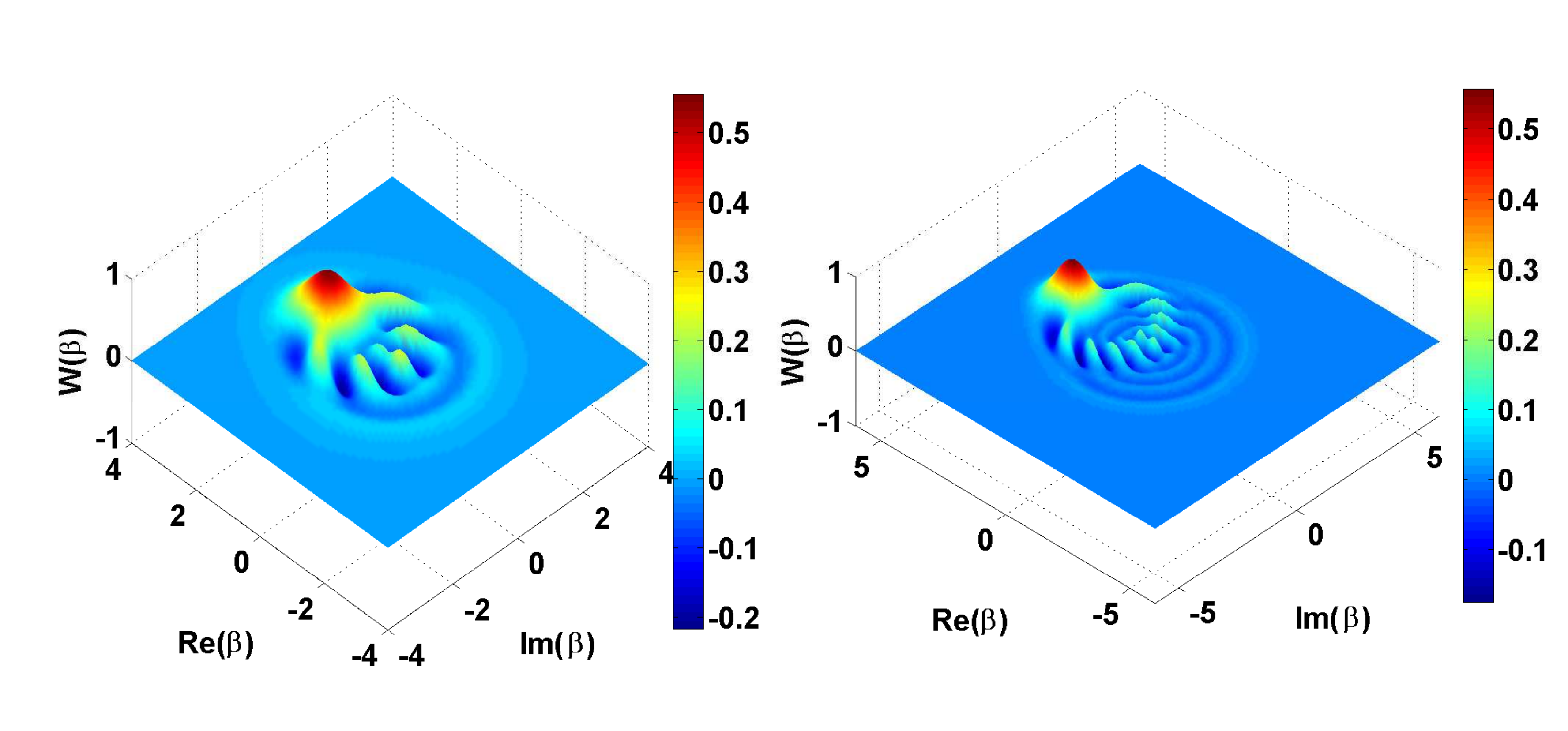}
\caption{Wigner function for the states $\ket{\psi(2,5)}$ (left) and $\ket{\psi(3,9)}$ (right) showing negative regions.}
\label{Wigner}
\end{figure}\\
For the state $\ket{\psi(\alpha,m)}$,
\begin{align}
W(\beta)=\frac{2}{\pi}\sum_{n=0}^\infty (-1)^n \left|e^{-i Im (\beta \alpha^*)}\langle \alpha-\beta|n\rangle-C_m^* D_{mn}(\beta)t\right|^2.
\end{align}
Here
\begin{align}
D_{mn}(\beta)&=\sqrt{\frac{n!}{m!}}e^{-|\beta|^2/2}\beta^{m-n}L_n^{(m-n)}(|\beta|^2),~~~m> n\\
D_{mn}(\beta)&=\sqrt{\frac{m!}{n!}}e^{-|\beta|^2/2}(-\beta^*)^{m-n}L_n^{(n-m)}(|\beta|^2), ~~~m\le n
\end{align}
where $L_n^{(k)}$ is the associated Laguerre polynomial.
\begin{figure}[h]
\centering
\includegraphics[width=9cm, height=5cm]{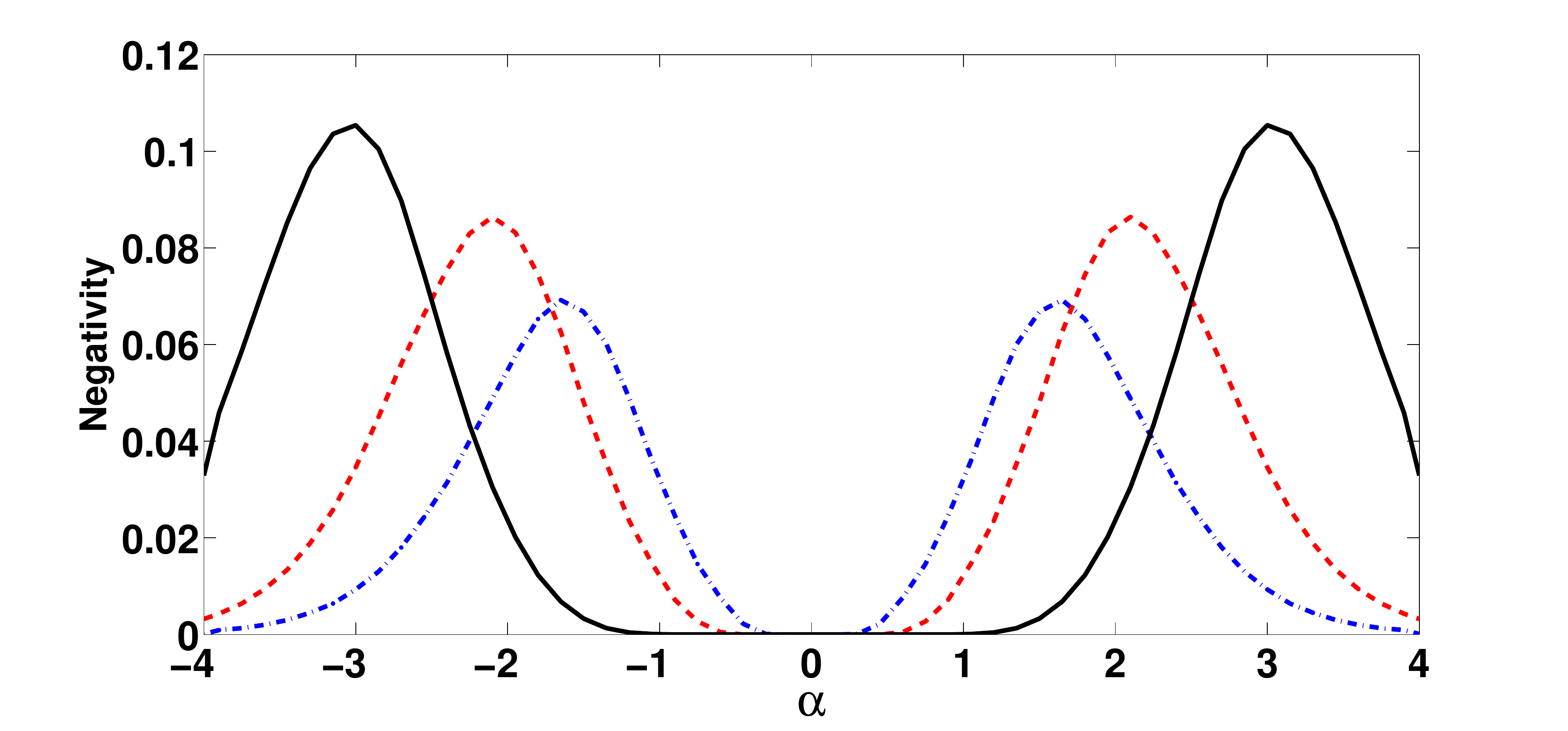}
\caption{Negativity of Wigner function as a function of $\alpha$ for $\hbs$ corresponds to $m=2$ (dot-dashed),4 (dashed) and 9(continuous).}
\label{Negativity}
\end{figure}\\
Wigner functions for $\ket{\psi(\alpha,m)}$ with $\alpha=2,m=5$ and $\alpha=3,m=9$ are shown in Fig. \ref{Wigner}. Both the states are nonclassical as their respective Wigner functions are negative in some regions of phase space. Interestingly, the number of peaks equals $m+1$. To quantify the nonclassicality,  the volume under negative portion of the Wigner function is shown in Fig. \ref{Negativity}.  It is readily inferred from the figure that the negativity is maximum when $\vert\alpha\vert^2 \approx m$.  This is consistent with the result that the overlap of $\hbs$ with the classical state $\ket{\alpha}$ reaches its minimum when $|\alpha|^2=m$.
\subsection{Beam splitter transformation}
Another way to quantify nonclassicality is the entanglement at output of a beam splitter. If one of the input states is a coherent state, it requires a nonclassical state at the second port to generate entanglement at the output \cite{Kim}. The amount of entanglement is a measure of the nonclassicality of the input state to the second port. The unitary transformation effected by a $50:50$ beam splitter is
\begin{align}
U=\left[\begin{array}{cc}
\frac{1}{\sqrt{2}} & \frac{i}{\sqrt{2}}\\
\frac{i}{\sqrt{2}} & \frac{1}{\sqrt{2}}
\end{array}\right].
\end{align}
The input bipartite state $\ket{\psi(\alpha,m)}\ket{\beta}$ transforms to
\begin{widetext}
\begin{align}
\ket{\psi_{AB}}=U\ket{\psi(\alpha,m)}\ket{\beta}=&\frac{1}{N_m}\left[ \sum_{r=0}^{\infty} e^{-|\tilde{\alpha}|^2/2}\frac{\tilde{\alpha}^r}{\sqrt{r!}}\sum_{s=0}^{\infty} e^{-|\tilde{\beta}|^2/2}\frac{\tilde{\beta}^s}{\sqrt{s!}}\ket{r,s}\right.\nonumber\\
&\left.-C_m \sum_{n=0}^{\infty} e^{-|\beta|^2/2}\frac{\beta^n}{\sqrt{n!}}\left(\sum_q^m\sum_{q'}^n\sqrt{\frac{(q+q')!(m+n-q-q')!}{m!n!}}\right.\right.\nonumber\\
&\left.\left.~~~~~~~~~~~\times\left(\frac{1}{\sqrt{2}}\right)^{n+q-q'}\left(\frac{i}{\sqrt{2}}\right)^{m-q+q'}\ket{q+q',m+n-q-q'}\right)\right],
\end{align}
\end{widetext}
where $\tilde{\alpha}=(\alpha+i\beta)/\sqrt{2}$ and $\tilde{\beta}=(i\alpha+\beta)/\sqrt{2}$. Here $\ket{\beta}$ is a coherent state.\\

\begin{figure}[h]
\centering
\includegraphics[width=9cm, height=6cm]{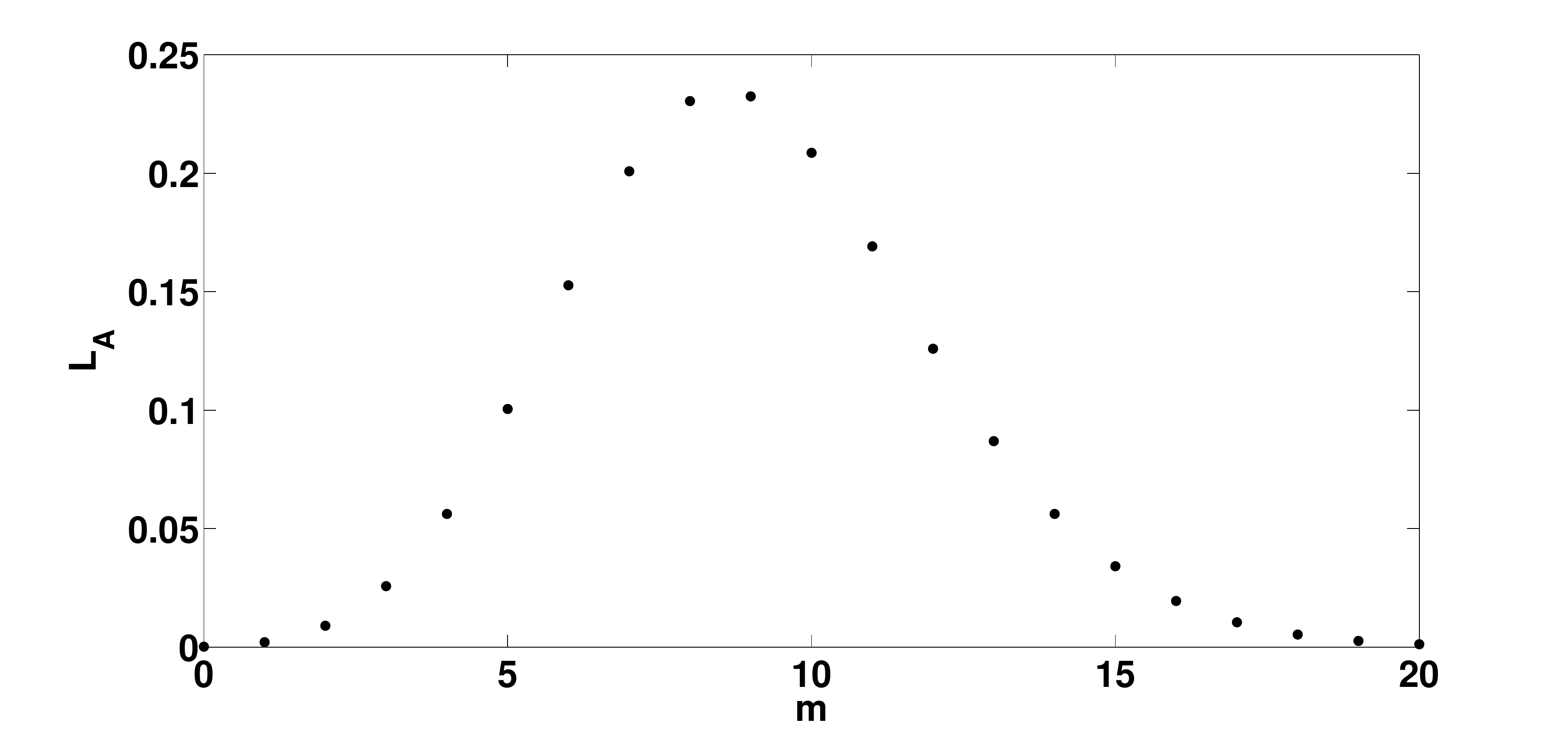}
\caption{Linear entropy $L_A$ for the state $\ket{\psi_{AB}}$ as a function of $m$ with $\alpha=\beta=3$.} 
\label{LE}
\end{figure}

Entanglement in the output of the beam splitter is quantified by the linear entropy of one of the reduced density operators of the bipartite output.
It is defined as \cite{Wei},
\begin{align}
L_A(\ket{\psi_{AB}})=1-Tr(\rho^2_A),
\end{align}
where $\rho_A=Tr_B(\ket{\psi_{AB}}\bra{\psi_{AB}})$. Here $A$ and $B$ are refer to the two output ports.
For pure bipartite states, linear entropy is a good measure of entanglement. Non-zero values of $L_A$ imply entanglement in $\ket{\psi_{AB}}$.
 Fig. \ref{LE} shows the entanglement in the output state $\ket{\psi_{AB}}$ for various values of $m$ with $\alpha=\beta=3$. It is clear from the figure that the state at the output ports of the beam splitter is entangled, which indicates that the input state is nonclassical. For large $m$, $\hbs\approx \ket{\alpha}$ and hence the entanglement of the output state is zero. Beam splitter generates maximally entangled state if $|\alpha|^2=m$ for a given $\alpha$ and independent of $\beta$. This is consistent with the observation made earlier in Section \ref{Nonclassical} that the nonclassicality of $\ket{\psi(\alpha,m)}$ is expected to be maximum if $|\alpha|^2=m$. 
\section{Generation of NSFS}\label{Generation}
In this section, a scheme to generate $\hbs$ is presented. This process involves driving a system of an atom and a cavity by two consecutive Hamiltonian evolutions. First, a multi-photon interaction drives the atom-cavity system. The resultant state is modified by driving the atomic system with a classical field. Details of the processes are indicated here.\\

\begin{figure}[h]
\centering
\includegraphics[scale=0.4]{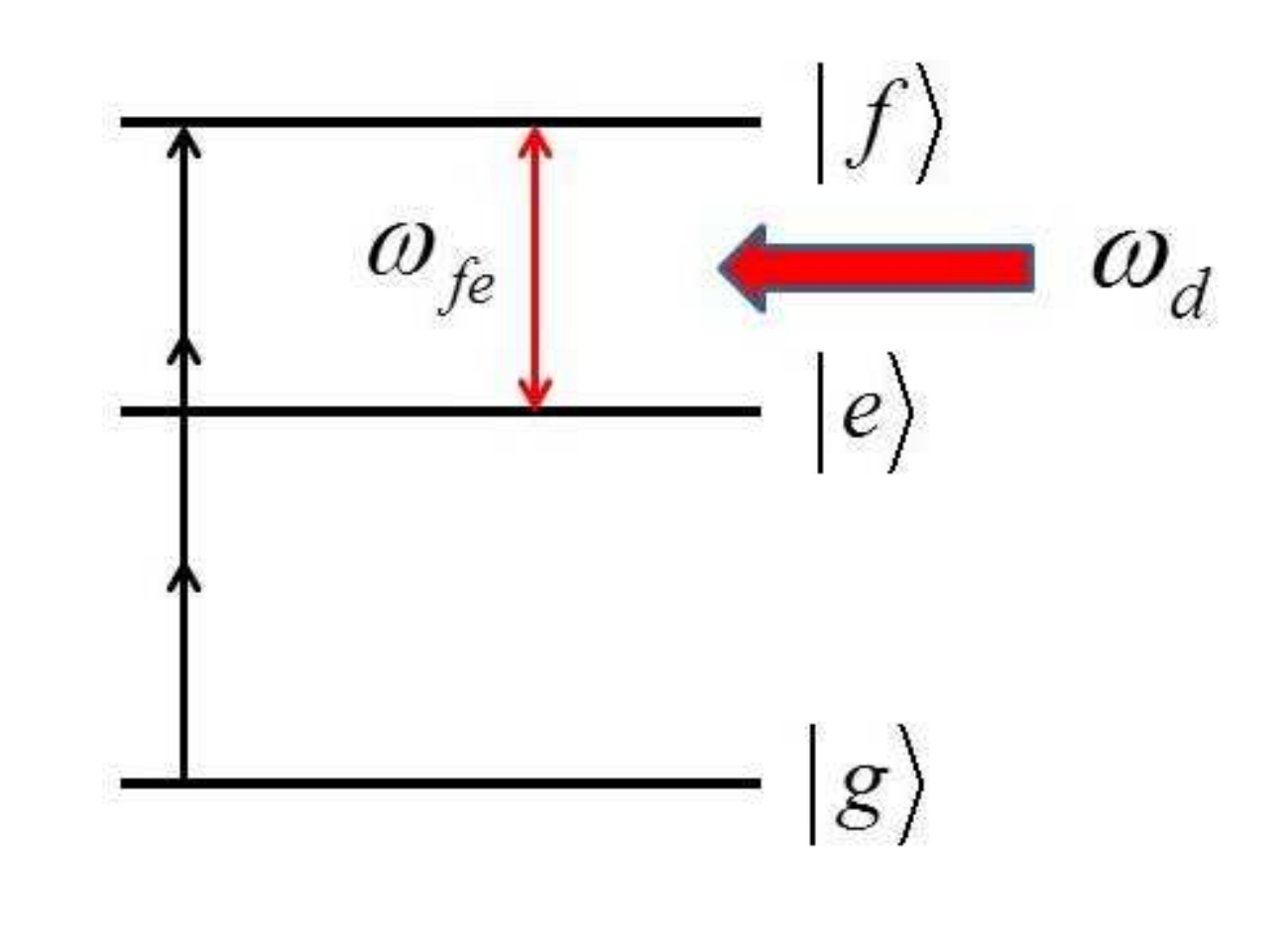}
\caption{Ladder type 3-level atom with energy levels $\ket{f},\ket{e}$ and $\ket{g}$ are labeled in decreasing order of their energies. }
\label{threelevelatom}
\end{figure}

Consider a $\Xi$ type 3-level atom interacting with a cavity field of resonance frequency $\omega_c$. The states of the 3-level atom are labeled by $\ket{f},\ket{e}$ and $\ket{g}$ in decreasing order of their energies as shown in Fig. \ref{threelevelatom}. Interaction between the field and the atom is given by the multi-photon Hamiltonian \cite{Zubairy}
\begin{align}
H_I=g(e^{i\phi}a^{\dagger m}\ket{g}\bra{f}+e^{-i\phi} a^m\ket{f}\bra{g}),
\end{align}
which connects the states $\ket{g}$ and $\ket{f}$.
The transition from $\ket{g}$ to $\ket{f}$ is achieved by annihilating $m$ photons of the cavity field and the transition from $\ket{f}$ to $\ket{g}$ creates $m$ photons of the cavity field.
The initial state of the atom-cavity is $\ket{0,f}$, where the cavity field is in its vacuum and the atom is in the state $\ket{f}$. The time evolved state at time $t_1$ is
\begin{align}
\ket{\psi(t_1)}=\cos(\sqrt{m!}gt_1)\ket{0,f}- i e^{im\phi}\sin(\sqrt{m!}gt_1)\ket{m,g}.
\end{align}
The atomic transition $\ket{e}\rightarrow \ket{f}$ comes in to the picture when the atom is driven by an external laser with Rabi frequency $\epsilon$. The corresponding Hamiltonian  is
\begin{align}
H_2=&\frac{1}{2}\hbar \omega_{fe}\sigma_z+\hbar \omega_c a^\dagger a+\hbar g(\sigma_+ a+\sigma_- a^\dagger)\nonumber\\
&+\hbar\epsilon( e^{-i\omega_{d}t}\sigma_-+e^{i\omega_{d}t}\sigma_+),
\end{align}
where $\sigma_z=\ket{f}\bra{f}-\ket{e}\bra{e}$, $\sigma_+=\ket{f}\bra{e}$ and $\sigma_-=\ket{e}\bra{f}$.  The transition frequency between the states $\ket{e}$ and $\ket{f}$ is $\omega_{fe}$. We assume $\omega_c=\omega_d$.
In the limit of large detuning ($\Delta=\omega_{fe}-\omega_c$) and in the rotating frame defined by $e^{-i\omega_{d}(\sigma_z+a^\dagger a)}$, the effective Hamiltonian corresponding to $H_2$ is \cite{Gerry2}
\begin{align}
H_{eff}=\hbar\chi [\ket{f}\bra{f}+(a^\dagger a+\lambda (a^\dagger+ a)+\lambda^2)\sigma_z],
\end{align}
where $\chi=g/(\omega_{fe}-\omega_c)$ and $\lambda=\epsilon/g$. If the state of the atom is $\ket{g}$, then the cavity field is unaffected. If the atomic state is $\ket{e}$ or $\ket{f}$, the cavity evolution is non-trivial. Cavity driven is conditional depending on the atomic state. \\

Under the evolution governed by $H_{eff}$, $\ket{\psi(t_1)}$ evolves to
\begin{align}
\ket{\psi(t_1+t_2)}=&\cos(\sqrt{m!}gt_1)e^{i\lambda^2\sin^2\chi t_2}\ket{-\lambda(1-e^{i\chi t_2)}}\ket{f} \nonumber\\&-ie^{im\phi}\sin(\sqrt{m!}gt_1)\ket{m,g},
\end{align}
where $\ket{-\lambda(1-e^{i\chi t_2})}$ is a coherent state. Using a Ramsey pulse to drive the transitions
$\ket{f}\rightarrow \frac{1}{\sqrt{2}}(\ket{f}+\ket{g})$ and $ 
\ket{g}\rightarrow \frac{1}{\sqrt{2}}(\ket{f}-\ket{g})$, 
the state $\ket{\psi(t_1+t_2)}$ transforms to
\begin{align}
\ket{\psi(t_1+t_2)}=(b\ket{\alpha}-c\ket{m})\ket{f}+(b\ket{\alpha}+c\ket{m})\ket{g},
\end{align}
where $\alpha=-\lambda(1-e^{i\chi t_2})$, $b=\frac{1}{\sqrt{2}}(\cos(\sqrt{m!}gt_1)e^{i\lambda^2\sin^2\chi t_2})$ and $c=\frac{1}{\sqrt{2}}ie^{im\phi}\sin(\sqrt{m!}gt_1)$. The resultant state contains the required field state $\hbs$ as a component. It is in post-selection of the atomic state, the cavity field is projected onto the required state. If the atom is detected in the state $\ket{f}$, the state of the field collapses to $(b\ket{\alpha}-c\ket{m})$ which is the target state $\hbs$ for proper choices of $b$ and $c$. 
\section{NSFS for quantum information processing}\label{QIP}
The elementary unit for quantum computation and information processing is a qubit. Any two orthogonal quantum states of a system form a qubit. In the context of cavities, the vacuum state and the single photon state form a qubit. Various quantum protocols such as  quantum teleportation \cite{Lombardi, Song}, quantum state transfer \cite{Meher}, entanglement generation\cite{Tan}, etc. have been implemented in this system.  SPACS has also been suggested for implementing various quantum information protocols \cite{Pinheiro, Wang2}. Performance of any quantum information processing system is affected by external factors such as dissipation, decoherence, etc. It has always been of interest to use qubits which are resilient against such processes.\\

 One way of assessing the robustness of the qubit evolving under a dissipative channel, is to calculate the overlap between its initial and final states. We consider four families of qubits using cavity field. One of the states of the qubit is the vacuum state. The other state can be a single photon state $\ket{1}$, a coherent state $\ket{\alpha}$, a SPACS $\ket{\alpha,1}$ or a NSFS $\ket{\psi(\alpha,0)}$. Of these choices, the one with $\ket{\alpha}$ is not ideal qubit as the states $\ket{0}$ and $\ket{\alpha}$ are not orthogonal.  A phenomenological model of damping in a quantum channel is considered to study the effect of dissipation on these families of qubits \cite{Gerry}. The fidelity, which is the overlap of evolved state and the initial state, is
\begin{align}
F_{\ket{\phi}}=\left| \langle \phi(0)\vert \phi(t)\rangle\right|^2.
\end{align}
where $\ket{\phi(t)}\propto e^{-\gamma a^\dagger a t/2}\ket{\phi(0)}$. The amplitude damping rate is $\gamma$. The respective fidelities for the states mentioned above are
\begin{align}
F_{\ket{1}}&=e^{-\gamma t},\nonumber\\
F_{\ket{\alpha}}&=e^{-|\alpha-\alpha_d|^2},\nonumber\\
F_{\ket{\psi(\alpha,0)}}&=\frac{\left|e^{-\frac{1}{2}|\alpha-\alpha_d|^2}-e^{-\frac{1}{2}(|\alpha|^2-|\alpha_d|^2)}\right|^2}{(1-e^{-|\alpha|^2})(1-e^{-|\alpha_d|^2})},\nonumber\\
F_{\ket{\alpha,1}}&=e^{\frac{1}{2}|\alpha-\tilde{\alpha}_d|^2}e^{\frac{1}{2}(|\alpha|^2-|\tilde{\alpha}_d|^2)}e^{-|\alpha-\alpha_d|^2}e^{-(|\alpha|^2-|\alpha_d|^2)}\nonumber\\
 &~~~~~~~~\times\frac{(1+\alpha^*\alpha_d)^2}{(1+|\alpha|^2)(1+\alpha^*\tilde{\alpha}_d)},
\end{align}
where $\alpha_d=\alpha e^{-\gamma t/2}$ and $\tilde{\alpha}_d= \alpha e^{-\gamma t}$. The suffix indicates the state whose fidelity is being studied.\\

\begin{figure}[h]
\centering
\includegraphics[width=9cm, height=5cm]{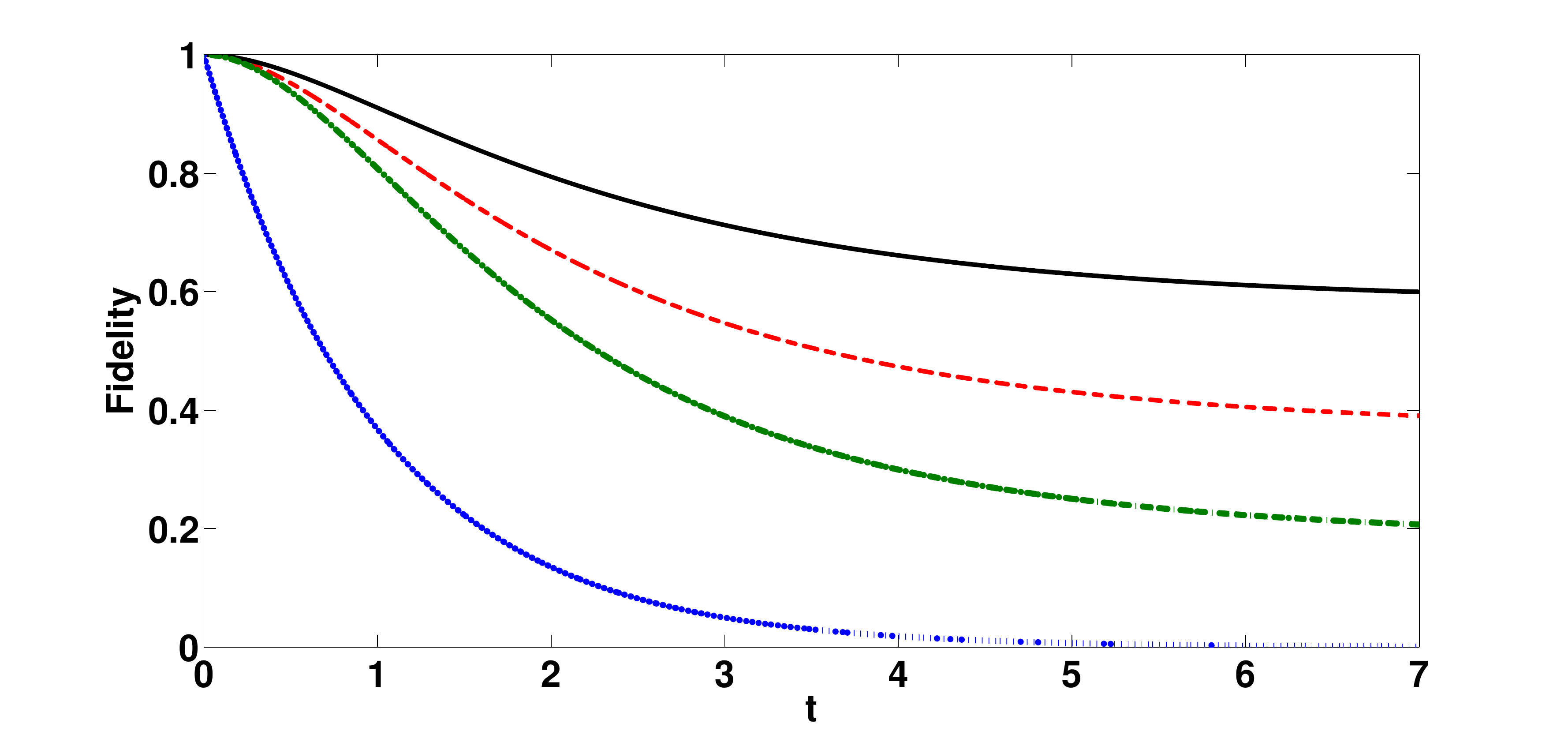}
\caption{Fidelities $F_{\ket{\psi(\alpha,0)}}$(continuous), $F_{\ket{\alpha}}$(dashed), $F_{\ket{\alpha,1}}$ (dot-dashed) and $F_{\ket{1}}$(dotted)   as a function of time. All curve corresponds to $|\alpha|=1$.   }
\label{Qubit}
\end{figure}

Fidelities for these various quantum states are shown in Fig. \ref{Qubit}, assuming $\gamma=0.01$. It is to be noted that the state $\ket{\psi(\alpha,0)}$ is robust against damping in a quantum channel as its fidelity decays slower than the other states. This makes the state  suitable for quantum information processing.\\ 

Secure communication protocols such as BB84 are being carried out using the coherent states. The drawback in using the coherent states is that the eavesdropper can extract some photons, which may lead to loss of information. In order to prevent photon loss, Mundarain \textit{et. al} have shown that SPACS can be used for two-way quantum key distribution \cite{Miranda}. Their protocol guarantees secure key distribution between two parties A and B even in the presence of a third party E who eavesdrops with a beam splitter. In order to quantify the information gathered by E, the respective probabilities of detecting at least one photon by B and E are calculated. If $\ket{0}$ and SPACS  are used, then these probabilities are
\begin{align}
P_{B}(n\geq 1)=1-\frac{e^{-|\alpha|^2 \cos^2\theta}}{1+|\alpha|^2}\left(\sin^2\theta+|\alpha|^2\sin^4\theta\right),
\end{align}
and 
\begin{align}
P_{E}(n\geq 1)=1-\frac{e^{-|\alpha|^2 \sin^2\theta}}{1+|\alpha|^2}\left(\cos^2\theta+|\alpha|^2\cos^4\theta\right),
\end{align}
 where $\theta$ is an angular parameter that characterize the beam splitter. It is clear that for small $\theta$ and  $\alpha$, information loss is less. \\
 
\begin{figure}[h]
\centering
\includegraphics[width=9cm, height=5cm]{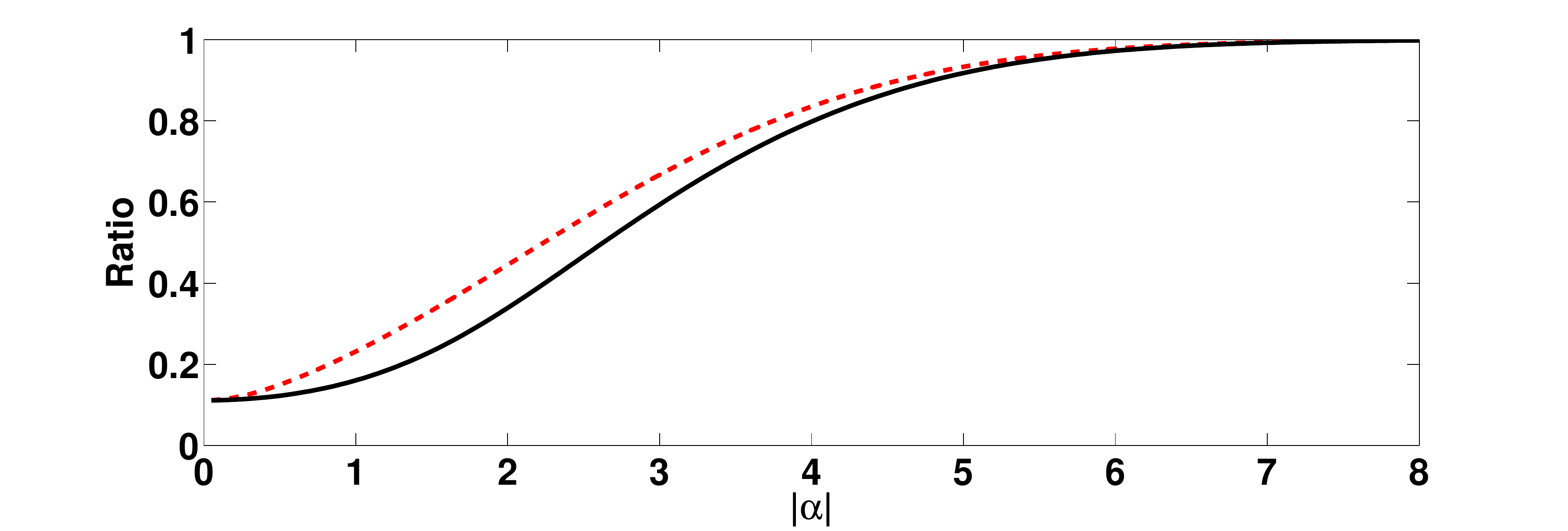}
\caption{ Ratios $R_{\ket{\psi(\alpha,0)}}$ (continuous line) and $R_{\ket{\alpha,1}}$ (dashed line) as a function of $|\alpha|$. Here $\cos^2\theta=0.9$.      }
\label{Ratio}
\end{figure}

The state $\ket{\psi(\alpha,0)}$ is more robust against dissipative losses in a quantum channel. This may endow $\ket{\psi(\alpha,0)}$ as a better choice than SPACS. The respective  probabilities of detecting at least one photon by B and E when $\ket{0}$ and $\ket{\psi(\alpha,0)}$  are the input to the beam splitter are 
\begin{align}
\tilde{P}_{B}(n\geq 1)=1-\frac{e^{-|\alpha|^2 \cos^2\theta}-e^{-|\alpha|^2}}{1-e^{-|\alpha|^2}},
\end{align}    
and
\begin{align}
\tilde{P}_{E}(n\geq 1)=1-\frac{e^{-|\alpha|^2 \sin^2\theta}-e^{-|\alpha|^2}}{1-e^{-|\alpha|^2}}.
\end{align}
In order to compare the advantage of $\ket{\psi(\alpha,0)}$ over SPACS, the ratios of detection probabilities are considered. These are defined as 
\begin{align}
R_{\ket{\psi(\alpha,0)}}=\frac{\tilde{P}_{E}(n\geq 1) }{\tilde{P}_{B}(n\geq 1)},
\end{align} 
and
\begin{align}
R_{\ket{\alpha,1}}=\frac{P_{E}(n\geq 1) }{P_{B}(n\geq 1)}.
\end{align} 
These two ratios are shown in Fig. \ref{Ratio}. The ratio $R_{\ket{\psi(\alpha,0)}}$ is always less than $R_{\ket{\alpha,1}}$ indicates that the diversion of photons by eavesdropper E \textit{via} beam splitter transformation is less probable if one uses the state $\ket{\psi(\alpha,0)}$ instead of SPACS. This shows $\ket{\psi(\alpha,0)}$ performs better than SPACS. 
\section{Generation of Schr$\ddot{\text{O}}$dinger cat state from NSFS }\label{Catstate}
Generation of the even and odd coherent states from $\hbs$ is discussed. It is possible to realize the aforementioned states, for instance, from coherent states in cavities \cite{Brune}, mixing squeezed vacuum and vacuum in beam splitter \cite{Wolinsky, Dakna, Scott}, etc. The procedure for the generation of the odd/even coherent states from  $\hbs$ is similar to that for the generation of these states from the coherent state \cite{Brune}. The system consists of a $\Xi$ type three-level atom as described in Section \ref{Generation} and a cavity with resonance frequency $\omega_c$. Consider the initial state
\begin{align}
\ket{\Psi(0)}=\frac{1}{\sqrt{2}}(\ket{e}+\ket{g})\hbs.
\end{align}
A dispersive atom-field coupling is assumed so that the effective Hamiltonian is
\begin{align}
H_{eff}=-\hbar \chi a^\dagger a \ket{e}\bra{e}.
\end{align}
Time evolved state under $H_{eff}$ is
\begin{align}
\ket{\Psi(t)}&=e^{-iH_{eff}t/\hbar}\ket{\Psi(0)},\nonumber\\
&=\frac{1}{\sqrt{2}}(\ket{e}\ket{\psi(e^{i\chi t}\alpha,m)}+\ket{g}\ket{\alpha,m)}),
\end{align} 
A Ramsey pulse is applied that transforms $\ket{e}\rightarrow (\ket{e}+\ket{g})/\sqrt{2}$ and $\ket{g}\rightarrow (\ket{g}-\ket{e})/\sqrt{2}$. The evolved state at time $t=\pi/\chi$ is
\begin{align}
\ket{\Psi(\pi/\chi)}=&\frac{1}{2}[\ket{e}(\ket{\psi(-\alpha,m)}-\ket{\psi(\alpha,m)})\nonumber\\
&+\ket{g}(\ket{\psi(-\alpha,m)}+\ket{\psi(\alpha,m)})].
\end{align}
It is easy to recognize that if $m$ is even and the atom is detected in the state $\ket{e}$, the field state collapses to the odd coherent state. If $m$ is odd and the atom is detected in $\ket{g}$, the field state is projected on to the even coherent state. 
\section{Summary}\label{Summary}
Coherent states are special as they exhibit minimum uncertainty in the quadratures, Poissonian statistics, factorizability of coherence functions, etc. All these features which make these states the most classical among  the quantum states arise due to the specific choice of superposition coefficients. Number state filtering from a coherent state leads to a nonclassical state. These states are realizable in the multi-photon interaction between a three-level atom and a single mode cavity field in a coherent state.  The number state $\ket{m}$ that has the maximum weightage in a coherent state $\ket{\alpha}$ corresponds to $|\alpha|^2\approx  m$.  This is the condition for maximal nonclassicality for a number state filtered coherent state. The overlap between the coherent state and NSFS is minimum when this condition is satisfied. The negativity of the Wigner function is maximum under these conditions. Also, the entanglement potential of these states is highest when $ |\alpha|^2 \approx m$. If $m$ deviates from $|\alpha|^2$, the resultant state is still nonclassical as shown by the sub-Poissonian. This property enables these states to perform slightly better than the coherent states if used for phase measurements in an interferometric setup.      Surprisingly,  NSFS is super-Poissonian when $|\alpha|^2\approx m$.  Emergence of this super-Poissonian statistics is understood based on the fact that NSFS is  a  superposition of suitably truncated coherent states which are sub-Poissonian.   The super-Poissonian statistics emerges in an analogous manner as in suitable  superposition of two number states which are also sub-Poissonian.     \\

  Further, vacuum state filtered coherent states are resilient against the effects of dissipation than the photon-added coherent states. This makes NSFS more suitable for quantum information processing. For instance, a robust qubit can be realized using the vacuum and NSFS as the computational basis states. The vacuum state filtered coherent state performs better than the single photon-added coherent state in two-way quantum key distribution where eavesdropping is carried out using a beam splitter. 

\section*{Acknowledgement} 
One of the authors (NM) acknowledges the research fellowship from the Department of Atomic Energy, Government of India.      
\bibliographystyle{apsrev4-1}
\bibliography{NumberStateFilteredCoherentState}
\end{document}